\documentclass[12pt,a4paper]{article}
\usepackage{amsmath,amssymb}
\usepackage{graphicx}
\begin{document}
\title{
A diagrammer's note on DC conductivity of anisotropic Fermi liquids for beginners: 
Maebashi-Fukuyama formula and Taylor formula
}

\author{
O. Narikiyo
\footnote{
Department of Physics, 
Kyushu University, 
Fukuoka 819-0395, 
Japan}
}

\date{
(August 23, 2016)
}

\maketitle
\begin{abstract}
Formal but exact DC conductivity formulae 
for anisotropic Fermi liquids are reviewed. 
One is the Maebashi-Fukuyama formula based on the Fermi-surface harmonics. 
The other is the Taylor formula based on the scattering eigenfunction. 
In comparison to these two formulae 
the current-vertex-correction in the fluctuation-exchange approximation 
is shown to be a bad vision caused by an inconsistent approximation. 
\end{abstract}
\vskip 20pt

\section{Introduction}

The DC conductivity formula obtained by the Fermi-liquid theory is shown 
to be equivalent to the one obtained by the linearized Boltzmann equation 
in the textbook~\cite{AGD}. 
This proof is done for the case of spherical Fermi surface. 
Its extension to the case of anisotropic Fermi surface 
is done by Maebashi and Fukuyama~\cite{MF1}. 
The Maebashi-Fukuyama formula is formal but exact. 

On the other hand, 
Taylor~\cite{Taylor} derived a formal but exact solution 
of the linearized Boltzmann equation. 

These two works employ opposite strategies. 

The Maebashi and Fukuyama formula is on the basis of the Fermi-surface harmonics 
which is the polynomial of the velocity. 
Then the interaction is expanded in terms of the velocity. 

The Taylor formula is on the basis of the scattering eigenfunction. 
In this formula the interaction is diagonalized 
but the velocity is expanded in terms of the scattering eigenfunction. 

By these two formulae 
we know the exact forms of the DC conductivity of anisotropic Fermi liquids. 
Consequently it becomes evident 
that there is no room for the current-vertex-correction 
discussed in the fluctuation-exchange (FLEX) approximation~\cite{FLEX}. 

\section{Maebashi-Fukuyama formula}

\noindent
{\large\bf 2-1 Fermi-surface harmonics expansion}

\vskip 10pt

We start from the Fermi-liquid formula~\cite{MF1} for the DC conductivity 
\begin{equation}
\sigma_{xx} = 2 e^2 \sum_{\bf p} \sum_{\bf p'} 
v_x({\bf p}) \big(K''\big)^{-1}_{{\bf p}{\bf p'}} \ v_x({\bf p'})
\Big( - { \partial f_{\bf p} \over \partial \epsilon_{\bf p} } \Big), 
\label{FL} 
\end{equation}
which is (2.34) of \cite{MF1} and 
equivalent to the result of the linearized Boltzmann equation, 
(58) of \cite{rev1}.  

Using the Fermi-surface harmonics~\cite{Allen,EG}, $\psi_L({\bf p})$, 
(\ref{FL}) is written as 
\begin{equation}
\sigma_{xx} = 2 e^2 \sum_{\bf p} \sum_{\bf p'} 
\psi_1({\bf p}) 
\sum_L \sum_{L'} \big(K''\big)^{-1}_{LL'} \psi_L({\bf p}) \psi_{L'}({\bf p'}) 
\psi_1({\bf p'}) 
\langle v_x^2 \rangle 
\Big( - { \partial f_{\bf p} \over \partial \epsilon_{\bf p} } \Big), 
\label{anisoFL}
\end{equation}
where $\psi_1({\bf p}) = v_x({\bf p}) / \sqrt{\langle v_x^2 \rangle}$ and 
\begin{equation}
\big(K''\big)^{-1}_{{\bf p}{\bf p'}} = 
\sum_L \sum_{L'} \big(K''\big)^{-1}_{LL'} \psi_L({\bf p}) \psi_{L'}({\bf p'}).  
\end{equation}
The normalization factor is defined as 
\begin{equation}
\langle v_x^2 \rangle = {1 \over N(\epsilon)} 
\sum_{\bf p} v_x({\bf p})^2 \ \delta\big( \epsilon - \epsilon_{\bf p} \big), 
\end{equation}
with 
\begin{equation}
N(\epsilon) = \sum_{\bf p} \delta\big( \epsilon - \epsilon_{\bf p} \big). 
\end{equation}

By the orthonormal condition 
\begin{equation}
{1 \over N(\epsilon)} \sum_{\bf p} 
\psi_L({\bf p}) \psi_{L'}({\bf p}) \delta\big( \epsilon - \epsilon_{\bf p} \big) 
= \delta_{LL'}, 
\label{orthonormal}
\end{equation}
(\ref{anisoFL}) is written as 
\begin{equation}
\sigma_{xx} = 2 e^2 \int d\epsilon N(\epsilon) \int d\epsilon' N(\epsilon') 
\big(K''\big)^{-1}_{11} 
\langle v_x^2 \rangle 
\Big( - { \partial f \over \partial \epsilon } \Big). 
\end{equation}
Here we have used the relation 
\begin{equation}
\sum_{\bf p} = 
\int d\epsilon \sum_{\bf p} \delta\big( \epsilon - \epsilon_{\bf p} \big). 
\end{equation}

Introducing the memory-function matrix 
\begin{equation}
\big(M''\big)^{-1}_{11} =  
\int d\epsilon' N(\epsilon') \big(K''\big)^{-1}_{11}, 
\end{equation}
we obtain 
\begin{equation}
\sigma_{xx} = 2 e^2 \int d\epsilon N(\epsilon) 
\big(M''\big)^{-1}_{11} 
\langle v_x^2 \rangle 
\Big( - { \partial f \over \partial \epsilon } \Big). 
\label{M}
\end{equation}

Finally 
(\ref{M}) is written as 
\begin{equation}
\sigma_{xx} = 2 e^2 \sum_{\bf p} 
v_x({\bf p}) 
\big(M''\big)^{-1}_{11} 
v_x({\bf p}) 
\Big( - { \partial f_{\bf p} \over \partial \epsilon_{\bf p} } \Big). 
\end{equation}

In conclusion 
the DC conductivity is determined by the memory-function matrix~\cite{MF1}. 

The microscopic calculation of the memory-function matrix 
based on the Fermi-liquid theory 
is carried out by Maebashi and Fukuyama~\cite{MF1,MF2}. 
To obtain a finite DC conductivity 
we have to take the contribution of the Umklapp process 
into the calculation of $M''$. 
Next we have to perform the matrix inversion from $M''$ to $(M'')^{-1}$. 

\vskip 20pt

\noindent
{\large\bf 2-2 Collision term}

\vskip 10pt

In the previous subsection 
it is clarified that the DC conductivity is determined 
by the memory-function matrix. 
In the Boltzmann equation it appears in the collision term. 
Thus we investigate the collision term after Allen~\cite{Allen} 
to obtain the feeling of the matrix structure. 

The collision term $I_{\bf p}$ for $n_{\bf p}$ is given by 
\begin{equation}
I_{\bf p} = \sum_{\bf p'} K''_{{\bf p}{\bf p'}} 
\Big\{ n_{\bf p'} - n_{\bf p} \Big\}, 
\label{C_p}
\end{equation}
as (47) of \cite{rev1}. 
Using the Fermi-surface harmonics (\ref{C_p}) is written as 
\begin{equation}
I_{\bf p} = \sum_{\bf p'} \sum_L \sum_{L'} 
K''_{LL'} \psi_L({\bf p}) \psi_{L'}({\bf p'}) 
\sum_{L''} n_{L''} 
\Big\{ \psi_{L''}({\bf p'}) - \psi_{L''}({\bf p}) \Big\}. 
\label{C_L}
\end{equation}
By the orthonormal condition (\ref{orthonormal}) 
and $ \psi_0({\bf p}) = 1$, 
(\ref{C_L}) is written as 
\begin{equation}
I_{\bf p} = \int d\epsilon' N(\epsilon') \sum_L \sum_{L'} \sum_{L''} 
K''_{LL'} \ n_{L''} \ \psi_L({\bf p}) 
\Big\{ \delta_{L'L''} - \delta_{L'0} \psi_{L''}({\bf p}) \Big\}. 
\label{C_L'}
\end{equation}
Introducing the memory-function matrix 
\begin{equation}
M''_{LL'} =  
\int d\epsilon' N(\epsilon') K''_{LL'}, 
\end{equation}
(\ref{C_L'}) is written as 
\begin{equation}
I_{\bf p} = \sum_L \psi_L({\bf p}) 
\Big\{ 
\sum_{L'} M''_{LL'} \ n_{L'} - 
\sum_{L''}\sum_{L'} C_{L''L'L} \ M''_{L''0} \ n_{L'}  
\Big\}, 
\label{C_L''}
\end{equation}
where 
$ C_{LL'L''} $ is the Clebsch-Gordan coefficient defined by 
\begin{equation}
\psi_{L}({\bf p}) \psi_{L'}({\bf p})
= \sum_{L''} C_{LL'L''} \ \psi_{L''}({\bf p}). 
\label{CG} 
\end{equation}
Thus we obtain the collision term $I_L$ for $n_L$ as 
\begin{equation}
I_L = 
\sum_{L'} M''_{LL'} \ n_{L'} - 
\sum_{L''}\sum_{L'} C_{L''L'L} \ M''_{L''0} \ n_{L'}. 
\end{equation}
For anisotropic Fermi surfaces 
$n_L$ couples to the other $n_{L'}$s via the memory-function matrix 
so that we have to perform the matrix inversion, 
$ M'' \rightarrow (M'')^{-1}$, 
to obtain the conductivity. 

The case of the spherical Fermi surface is special 
and we do not have to perform the matrix inversion. 
By the symmetry $ K''_{{\bf p}{\bf p'}}$ only depends 
on the angle between ${\bf p}$ and ${\bf p'}$ 
so that the memory-function matrix becomes a diagonal matrix. 
Since $ C_{0L'L''} = \delta_{L'L''} $ 
which is obtained from (\ref{CG}) by using $ \psi_0({\bf p}) = 1$, 
we obtain 
\begin{equation}
I_L = 4\pi \Big\{ w_L - w_0 \Big\} n_L, 
\end{equation}
where 
we put $ M''_{LL'} = 4\pi w_L \delta_{LL'} $. 
This is the result shown in \cite{Peierls}. 
The DC conductivity is determined by $ n_1 $ 
and the transport life-time $\tau$ is given by 
\begin{equation}
{1 \over \tau} = 4\pi \Big\{ w_0 - w_1 \Big\}. 
\label{Peierls} 
\end{equation}

\vskip 20pt

\noindent
{\large\bf 2-3 Vector mean free path}

\vskip 10pt

The use of the vector mean free path ${\bf \Lambda}$ 
is convenient~\cite{Taylor,Ziman} 
for the discussion of the symmetry property of the conductivity tensor. 
In the following all the vectors are defined on the Fermi surface. 
The conductivity tensor ${\boldsymbol \sigma}$ is expressed 
in the dyadic form as~\cite{MF1,MF2,Taylor,Ziman} 
\begin{equation}
{\boldsymbol \sigma} = 
2 e^2 \sum_{\bf p} {\bf v}_{\bf p} \ {\bf \Lambda}_{\bf p}
\Big( - { \partial f_{\bf p} \over \partial \epsilon_{\bf p} } \Big), 
\label{dyadic} 
\end{equation}
where 
\begin{equation}
{\bf \Lambda}_{\bf p} = \sum_{\bf p'} 
\big(K''\big)^{-1}_{{\bf p}{\bf p'}} \ {\bf v}_{\bf p'}, 
\label{Lambda} 
\end{equation}
with 
$\displaystyle {\bf v}_{\bf p} 
= \big( v_x({\bf p}), v_y({\bf p}), v_z({\bf p}) \big)$. 

For systems with cubic symmetry 
the conductivity tensor reduces to a scalar $\sigma$~\cite{Taylor}. 
This means the integrand in (\ref{dyadic}) is proportional to 
the scalar product $\displaystyle {\bf v}_{\bf p} \cdot {\bf v}_{\bf p}$. 
Thus introducing the anisotropic life time $\tau_{\bf p}$, 
the vector mean free path is written as 
\begin{equation}
{\bf \Lambda}_{\bf p} = \tau_{\bf p} \ {\bf v}_{\bf p}. 
\label{cubic} 
\end{equation}
Namely, the direction of ${\bf \Lambda}_{\bf p}$ 
is the same as that of ${\bf v}_{\bf p}$. 
Substituting (\ref{cubic}) into (\ref{Lambda}) we obtain 
\begin{equation}
\tau_{\bf p} = \sum_{\bf p'} 
\big(K''\big)^{-1}_{{\bf p}{\bf p'}} \ 
{ {\bf v}_{\bf p} \cdot {\bf v}_{\bf p'} \over 
  {\bf v}_{\bf p} \cdot {\bf v}_{\bf p} }. 
\label{aniso-tau} 
\end{equation}
This relation (\ref{aniso-tau}) is a mere rewriting and has no practical use. 
Using this anisotropic life time the conductivity is expressed as 
\begin{equation}
\sigma_{xx} = 2 e^2 \sum_{\bf p} 
v_x({\bf p}) \ \tau_{\bf p} \ v_x({\bf p})
\Big( - { \partial f_{\bf p} \over \partial \epsilon_{\bf p} } \Big). 
\label{simple} 
\end{equation}
The Fermi-liquid formula \cite{Yamada} by Yamada and Yosida 
corresponds to this case. 
The appearance of (\ref{simple}) is simple but it is none other than (\ref{FL}). 
Actually 
\begin{equation}
\sigma_{xx} = 2 e^2 \sum_{\bf p} 
{ {\bf v}_{\bf p} \cdot {\bf v}_{\bf p} \over 3 } \ \tau_{\bf p} 
\Big( - { \partial f_{\bf p} \over \partial \epsilon_{\bf p} } \Big) 
= 2 e^2 \sum_{\bf p} \sum_{\bf p'} 
\big(K''\big)^{-1}_{{\bf p}{\bf p'}} \ 
{ {\bf v}_{\bf p} \cdot {\bf v}_{\bf p'} \over 3 } 
\Big( - { \partial f_{\bf p} \over \partial \epsilon_{\bf p} } \Big), 
\end{equation}
in the case of cubic symmetry. 

For systems without cubic symmetry 
the proportional relation (\ref{cubic}) does not hold~\cite{Taylor,Ziman} 
so that we have to use the Maebashi-Fukuyama formula discussed above. 

For systems with spherical symmetry 
$\tau_{\bf p}$ reduces to a constant $\tau$. 
In this case the inverse relation of (\ref{Lambda}) 
\begin{equation}
{\bf v}_{\bf p} = \sum_{\bf p'} 
K''_{{\bf p}{\bf p'}} \ {\bf \Lambda}_{\bf p'}, 
\end{equation}
reduces to 
\begin{equation}
{\bf v}_{\bf p} = \tau \sum_{\bf p'} 
K''_{{\bf p}{\bf p'}} \ {\bf v}_{\bf p'}. 
\end{equation}
Thus we obtain 
\begin{equation}
{1 \over \tau} = \sum_{\bf p'} 
K''_{{\bf p}{\bf p'}} \ \big( {\bf {\hat p}} \cdot {\bf {\hat p'}} \big), 
\label{life-time} 
\end{equation} 
with 
$ {\bf v}_{\bf p} = {\bf p} / m $ and $ {\bf {\hat p}} = {\bf p}/|{\bf p}| $. 
Since 
\begin{equation}
K''_{{\bf p}{\bf p'}} = 
{1 \over \tau_0} \delta_{{\bf p}{\bf p'}} - W_{{\bf p}{\bf p'}}, 
\end{equation}
and 
\begin{equation}
{1 \over \tau_0} = \sum_{\bf p'} W_{{\bf p}{\bf p'}}, 
\end{equation}
(\ref{life-time}) is written as 
\begin{equation}
{1 \over \tau} = \sum_{\bf p'} 
W_{{\bf p}{\bf p'}} \big( 1 - \cos\theta \big), 
\label{constant} 
\end{equation} 
with 
$ \cos\theta = {\bf {\hat p}} \cdot {\bf {\hat p'}} $. 
This expression (\ref{constant}) is none other than (\ref{Peierls}). 

\section{Taylor formula}

\noindent
{\large\bf 3-1 Setup}

\vskip 10pt

Taylor~\cite{Taylor} introduced the vector mean free path ${\bf \Lambda}$ 
and wrote down the DC conductivity tensor ${\boldsymbol \sigma}$ 
in the dyadic form as 
\begin{equation}
{\boldsymbol \sigma} = 
2 e^2 \sum_{\bf p} {\bf v}({\bf p}) \ {\bf \Lambda}({\bf p})
\Big( - { \partial f_{\bf p} \over \partial \epsilon_{\bf p} } \Big). 
\end{equation}
In the following all the vectors are defined on the Fermi surface. 
The vector mean free path is determined by 
\begin{equation}
{\bf v}({\bf p}) = \sum_{\bf p'} 
Q({\bf p},{\bf p'}) \big[ 
{\bf \Lambda}({\bf p}) - {\bf \Lambda}({\bf p'}) \big]. 
\label{vector}
\end{equation}
The Fermi-liquid theory works~\cite{MF1} 
on the calculation of the collision operator $Q$.  
Hereafter we only consider the $x$-component of (\ref{vector})  
\begin{equation}
v_x({\bf p}) = \sum_{\bf p'} 
Q({\bf p},{\bf p'}) \big[ 
\Lambda_x({\bf p}) - \Lambda_x({\bf p'}) \big]. 
\label{x-com}
\end{equation}

Introducing the inner product 
\begin{equation}
(a,b) \equiv \sum_{\bf p}
\Big( - { \partial f_{\bf p} \over \partial \epsilon_{\bf p} } \Big) 
a({\bf p})b({\bf p}), 
\end{equation}
the DC conductivity is compactly written as\footnote{
See, for example, 
Ashcroft and Mermin: 
{\it Solid State Physics} 
(1976) Chap. 16 - PROBLEM 4. 
} 
\begin{equation}
\sigma_{xx} = 2 e^2 \big( v_x, \Lambda_x \big). 
\label{DC}
\end{equation}

\vskip 20pt

\noindent
{\large\bf 3-2 Loose explanation}

\vskip 10pt

Loosely (\ref{x-com}) is written as 
\begin{equation}
v_x = \Big( {1 \over \tau} - {\hat Q} \Big) \Lambda_x, 
\label{matrix-x}
\end{equation}
where 
\begin{equation}
{\hat Q} \Lambda_x \equiv 
\sum_{\bf p'} Q({\bf p},{\bf p'}) \Lambda_x({\bf p'}), 
\end{equation}
and 
\begin{equation}
{1 \over \tau} \Lambda_x \equiv 
\sum_{\bf p'} {1 \over \tau({\bf p})} 
\delta_{{\bf p}{\bf p'}} \Lambda_x({\bf p'}), 
\end{equation}
with 
\begin{equation}
{1 \over \tau({\bf p})} \equiv 
\sum_{\bf p'} Q({\bf p},{\bf p'}). 
\end{equation}
The inverse of (\ref{matrix-x}) is 
\begin{equation}
\Lambda_x = \Big( {1 \over \tau} - {\hat Q} \Big)^{-1} v_x 
= \tau v_x + \tau {\hat Q} \tau v_x + \tau {\hat Q} \tau {\hat Q} \tau v_x 
+ \cdot\cdot\cdot. 
\label{inversion} 
\end{equation}
The perturbative expansion for the DC conductivity is shown in Fig. 1. 

\vskip 8mm
\begin{figure}[htbp]
\begin{center}
\includegraphics[width=12cm]{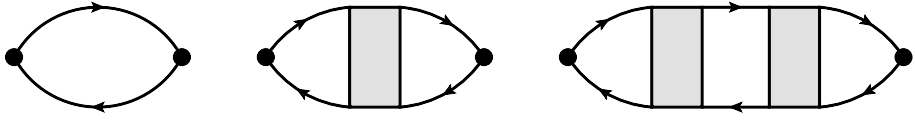}
\caption{
Diagrammatic representation of 
the first three inner products for the DC conductivity: 
$ (v_x,\tau v_x) $, $ (v_x,\tau {\hat Q} \tau v_x) $ and 
$ (v_x,\tau {\hat Q} \tau {\hat Q} \tau v_x) $. 
} 
\label{fig:1}
\end{center}
\end{figure}

If we can find the scattering eigenfunction $a_s({\bf p})$ 
which satisfies 
\begin{equation}
\tau({\bf p}) 
\sum_{\bf p'} Q({\bf p},{\bf p'}) a_s({\bf p'}) = \alpha_s a_s({\bf p}), 
\label{Q-alpha}
\end{equation}
with the eigenvalue $\alpha_s$, 
the inversion (\ref{inversion}) is readily accomplished. 
To proceed we expand the velocity in terms of the scattering eigenfunction as 
\begin{equation}
\tau({\bf p}) v_x({\bf p}) = \sum_s \beta_s a_s({\bf p}). 
\label{v-beta}
\end{equation}
Employing (\ref{Q-alpha}) and (\ref{v-beta}) in (\ref{inversion}) 
we obtain\footnote{
\begin{equation}
\tau {\hat Q} \tau v_x = \tau({\bf p}) \sum_{\bf p'} Q({\bf p},{\bf p'}) 
\sum_s \beta_s a_s({\bf p'})
= \sum_s \beta_s \alpha_s a_s({\bf p}), 
\nonumber
\end{equation}
and so on. 
}  
\begin{equation}
\Lambda_x({\bf p}) = \sum_s 
\big[ 1 + \alpha_s + \alpha_s^2 + \cdot\cdot\cdot \big]
\beta_s a_s({\bf p}), 
\label{Lambda-ab}
\end{equation}
To proceed further 
we have to establish the orthonormality of the scattering eigenfunction.

\vskip 20pt

\noindent
{\large\bf 3-3 Tight explanation}

\vskip 10pt

First we rewrite (\ref{Q-alpha}) as 
\begin{equation}
\sum_{\bf p'} \sqrt{\tau({\bf p})}
Q({\bf p},{\bf p'})\sqrt{\tau({\bf p'})}\  
{ a_s({\bf p'}) \over \sqrt{\tau({\bf p'})} }  
= \alpha_s \ 
{ a_s({\bf p}) \over \sqrt{\tau({\bf p})} }, 
\end{equation}
Since $ \sqrt{\tau({\bf p})}Q({\bf p},{\bf p'})\sqrt{\tau({\bf p'})} $ 
is a Hermitian matrix\footnote{
$Q({\bf p},{\bf p'})$ itself is a Hermitian matrix.}, 
the function $a_s({\bf p}) / \sqrt{\tau({\bf p})}$ 
constitutes an orthogonal basis. 
Thus the orthonormal condition is written as 
\begin{equation}
\Big( {a_s \over \sqrt{\tau}}, {a_{s'} \over \sqrt{\tau}} \Big)
= \Big( a_s, {a_{s'} \over \tau} \Big) = \delta_{ss'}. 
\label{orthonormal-a}
\end{equation}

Employing (\ref{v-beta}) and (\ref{Lambda-ab}) in (\ref{DC}) 
we obtain 
\begin{equation}
\sigma_{xx} = 2 e^2 \sum_{s'} \sum_s 
\beta_{s'} {1 \over 1 - \alpha_s } \beta_s 
\Big( {a_{s'} \over \tau}, a_s \Big). 
\end{equation}
Finally making use of the orthonormal condition (\ref{orthonormal-a}) 
we obtain the Taylor formula for the DC conductivity 
\begin{equation}
\sigma_{xx} = 2 e^2 \sum_s 
\beta_{s} {1 \over 1 - \alpha_s } \beta_s. 
\end{equation}
The repeated scatterings shown in Fig. 1 
result in the renormalization of the life time $\tau$. 

\section{Current-vertex-correction in FLEX approximation}

The purpose of this note is to clarify the exact expression 
for the DC conductivity of anisotropic Fermi liquids. 

Consequently it becomes evident that 
there is no room for the current-vertex-correction 
discussed in the FLEX approximation~\cite{FLEX}. 

Other deficiencies of the FLEX approximation 
have been also discussed previously~\cite{noFLEX1,noFLEX2}. 


\end{document}